\documentclass[12pt]{article}
\usepackage{amsmath,amssymb,amsthm,amssymb,graphicx}

\title{Totality of Subquantum Nonlocal Correlations}
\author{Andrei Khrennikov\\
International Center for Mathematical Modelling
\\in Physics and Cognitive Sciences,\\
Linnaeus University, S-35195, Sweden
}
\begin{document}

\maketitle

\abstract{In a series of previous papers we developed a purely
field model of microphenomena, so called prequantum classical
statistical field theory (PCSFT). This model not only reproduces
important probabilistic predictions of QM including correlations
for entangled systems, but it also gives a possibility to go
beyond quantum mechanics (QM), i.e., to make predictions of
phenomena which could be observed at the subquantum level. In this
paper we discuss one of such predictions -- existence of nonlocal
correlations between  prequantum random fields corresponding to
{\it all} quantum systems. (And by PCSFT quantum systems are
represented by classical Gaussian random fields and quantum
observables by quadratic forms of these fields.) The source of
these correlations is the common background field. Thus all
prequantum random fields are ``entangled'', but in the sense of
classical signal theory. On one hand, PCSFT demystifies quantum
nonlocality by reducing it to nonlocal classical correlations
based  on the common random background. On the other hand, it
demonstrates total generality of such correlations. They exist
even for distinguishable quantum systems in factorizable states
(by PCSFT terminology -- for Gaussian random fields with
covariance operators corresponding to factorizable  quantum
states).}

 \section{Introduction}

Tremendous development of quantum technologies provides new
intriguing possibilities for tests of foundations of QM and even
generates expectations to test predictions of prequantum models,
i.e., models describing microphenomena at the subquantum
 level and reproducing QM as emerging theory.
 In a series of previous papers \cite{KHR} I developed a purely
 field model of microphenomena, so called prequantum classical
statistical field theory (PCSFT). By this theory quantum systems
are represented by classical Gaussian random fields. Correlations
of quantum observables $A_1$ and $A_2$ on a composite system
$S=(S_1, S_2)$ are represented as correlations of quadratic forms,
$f_{A_1}(\phi_1), f_{A_2}(\phi_2),$ of components of the
prequantum random field $\omega \to \phi(\omega)=(\phi_1(\omega),
\phi_2(\omega))$ representing $S$ at the subquantum level. (Here
$\omega$ is a random parameter.)  PCSFT not only reproduces
important probabilistic predictions of QM including correlations
for entangled systems, but it gives a possibility to go beyond QM,
i.e., to make predictions on phenomena which could be observed at
the subquantum level. In particular, our prequantum model predicts
not only correlations of functionals (quadratic forms) of the
prequantum field, $\phi \to f(\phi),$ corresponding to quantum
observables, but even correlations between components of the
prequantum field corresponding to the subsystems $S_1$ and $S_2$
of $S.$ These correlations are always nonzero, even for random
fields corresponding to quantum systems in factorizable states,
$\Psi= \Psi_1 \otimes \Psi_2.$ We investigate this property of the
prequantum field in very detail and found that the situation is
very tricky from the probabilistic viewpoint. Although components
of the prequantum field are always correlated, all functionals of
this field corresponding to quantum observables for $S_1$ and
$S_2$ are never correlated (in the case of factorizable states).
Thus this effect, nonzero correlation between prequantum random
fields representing e.g. proton and electron which had been
prepared independently  e.g. at V\"axj\"o and Moscow, cannot be
found by using ``ordinary quantum observables''. New more delicate
measurement procedures have to be designed. In aforementioned
example we took proton and electron to emphasize that correlations
on the subquantum level have no direct relation to the quantum
entanglement for indistinguishable quantum systems, e.g., a pair
of electrons. Moreover, PCSFT predicts correlations even between
prequantum random fields corresponding to bosons and fermions
(even in the case of factorizable states). What is a source of
subquantum correlations? Why are prequantum random fields
corresponding to all quantum particles correlated?

PCSFT is heavily based on the assumption of the presence of a
sufficiently strong {\it background field,} cf. with stochastic electrodynamics. It is impossible to introduce a {\it
positively defined} covariance operator and hence to construct a
classical random field representation for quantum systems without
such a field. We can speculate that the common background field is
responsible for totality of correlations at the subquantum level.
The situation is very special from the probabilistic viewpoint.
For some quantum states, entangled states of QM, it is in
principle impossible to separate the background field from so to
say the intrinsic field of e.g. an electron. The latter field does
not exist as a classical random field (this  is again the problem
of positive definiteness of the covariance operator).  For
entangled states, the contribution of the background field can be
separated from the ``intrinsic contribution'' only on the level of
averages: the quantum average is obtained as the difference
between the average with respect the ``total prequantum signal''
and the average with respect to the background signal (a kind of
calibration procedure), see (\ref{YY1}), (\ref{YY2}).

In this situation of totality of mutual correlations it is natural
to consider a {\it prequantum  grand field}. Random fields
corresponding to quantum systems are simply random signals
generated by such a grand field. Hence, this work can be
considered as a step towards {\it classical unified field theory.}
However, this is a very preliminary step.

We start with a brief review of our previous results about
classical representation of quantum correlations, see \cite{KHR}
for details. Then we modify the previously developed formalism to
obtain a symmetric coupling between quantum and classical
covariances. Then we show that quadratic forms of Gaussian signals
corresponding to quantum systems in factorizable states are not
correlated. Finally, we present a criterium  to distinguish
prequantum random fields corresponding to entangled quantum
systems from fields corresponding to disentangled ones.

To simplify considerations, we will study quantum systems with
finite-dimensional Hilbert spaces. Moreover, we consider a
toy-model with the real Hilbert state space.

 \section{Classical representation of quantum correlations}

Take a Hilbert space $H$ as the space of states of classical
random fields. In classical signal theory $H=L_2({\bf R}^3).$ To
escape from mathematical problems we work in the
finite-dimensional case. However, in the appendix we work (on the
physical level of rigorousness) with $H=L_2({\bf R}^3).$

Consider a probability distribution $P$ on  $H$  having zero
average (it means that $\int_H (y, \phi) d P(\phi)=0$ for any $y
\in M$) and the covariance operator $D:$
\begin{equation}
\label{CO} (D y_1, y_2)=\int_H (y_1, \phi) (y_2, \phi) d P(\phi),
y_1, y_2 \in M.
\end{equation}
The $P$ can be considered as the probability distribution of an
$H$-valued random variable -- random  field (signal) (the
terminology which matches better the case $H=L_2({\bf R}^3)).$ We
remark that a covariance operator does not determine the random
signal uniquely. However, in the Gaussian case each $D$ determines
uniquely the Gaussian measure with zero mean value.

Let $H={\bf C}^n$ and $\phi(\omega)= (\phi_1(\omega),
...,\phi_n(\omega)),$ then zero average condition  is reduced to $
E\phi_k\equiv \int_\Omega \phi_k(\omega) dP(\omega)=0, k=1,..., n,
$  where $E$ is the operation of classical mathematical
expectation; the covariance matrix $D= (d_{kl}),$ where $
d_{kl}=E\phi_k \phi_l \equiv \int_\Omega \phi_k(\omega)
\phi_l(\omega) dP(\omega). $ We also recall that the dispersion of
the random variable $\phi$ is given by
$$
\sigma^2_\phi = E \Vert \phi(\omega) - E \phi(\omega)\Vert^2=
\sum_{k=1}^n  E\vert \phi_k(\omega) - E\phi_k(\omega)\vert^2.
$$
In the case of zero average we simply have
$$
\sigma^2_\phi = E \Vert \phi(\omega) \Vert^2= \sum_{k=1}^n  E\vert
\phi_k(\omega)\vert^2.
$$

\subsection{Operator representation of wave function of composite system}

In this section we show that the wave function of a composite
system has an operator representation which is useful in coupling
quantum and  classical correlations at the subquantum level, see
\cite{KHR}.

Let $H$ be a real Hilbert space. We denote the space of
self-adjoint operators acting in $H$ by the symbol ${\cal
L}_s(H).$ Since in this paper we consider only the finite
dimensional real case, this space can be realized as the space of
all symmetric matrices.

Let $H_1$ and $H_2$ be two real (finite dimensional) Hilbert
spaces. We put $H=  H_1 \otimes H_2.$ Any vector $\Psi \in H$  can
be represented in the form
$\Psi= \sum_{j=1}^m \psi_j \otimes \chi_j, \; \psi_j \in H_1,
\chi_j \in H_2,$
and it determines a linear operator from $H_2$ to $H_1$
$$
\widehat{\Psi} \phi= \sum_{j=1}^m (\phi, \chi_j) \psi_j, \; \phi
\in H_2.
$$ Of course, $\widehat{\Psi} \widehat{\Psi}^*: H_1 \to H_1$ and
$\widehat{\Psi}^* \widehat{\Psi}: H_2 \to H_2$ and these operators
are self-adjoint and positively defined.
Consider  operator  $\rho= \Psi \otimes \Psi: H_1\otimes H_2 \to H_1\otimes H_2$   and the  operators
$\rho^{(1)}\equiv \rm{Tr}_{H_2} \rho$ and $\rho^{(2)}\equiv
\rm{Tr}_{H_1} \rho.$ If the vector $\Psi$ is normalized by 1, then
$\rho$ is the density operator corresponding to the pure state
$\Psi$ and the operators $\rho^{(i)}\equiv \rho^{S_i}, i=1,2,$ are
the reduced density operators. These density operators describe
quantum states of subsystems $S_i, i=1,2,$ of a composite quantum
system $S=(S_1, S_2).$
For any $\Psi \in H_1\otimes H_2,$ the following equalities hold:
$$\rho^{(1)} = \widehat \Psi \widehat \Psi^* , \;
\rho^{(2)} = \widehat \Psi^* \widehat \Psi.$$

For any pair of  operators $\widehat{A}_j \in {\cal L}_s (H_j), j=
1,2, $ the following equality holds \cite{KHR}:
\begin{equation}
\label{Zuzu} \rm{Tr} \widehat{\Psi} \widehat{A}_2 \widehat{\Psi}^*
\widehat{A}_1= \langle \widehat{A}_1 \otimes \widehat{A}_2
\rangle_\Psi \equiv (\widehat{A}_1 \otimes \widehat{A}_2 \Psi,
\Psi).
\end{equation}
It will play a fundamental role in representation of quantum
correlations as classical correlations of quadratic forms of the
prequantum random field.

Let the state vectors of  systems $S_1$ and $S_2$ belong to
Hilbert spaces $H_1$ and $H_2,$ respectively. Then by  QM the
state vector $\Psi$ of the composite system $S=(S_1, S_2)$ belongs
to $H= H_1 \otimes H_2.$ We remark that the interpretation of the
state vector $\Psi \in H$ of a composite system is not as
straightforward as for a single system. It is known that, in
general, a pure state $\Psi$ of a composite system does not
determine pure states for its components. This viewpoint matches
well our approach. We shall interpret a normalized vector $\Psi
\in H$ not as the state vector of a concrete composite system
$S=(S_1, S_2),$ but as one of blocks of the covariance operator
for the prequantum random field $\omega \to \phi (\omega)= (\phi_1
(\omega), \phi_2 (\omega))$ describing $S=(S_1, S_2).$

\subsection{From classical to quantum correlations}
\label{EER}

Let $\phi_1$ and $\phi_2 $ be two random vectors,   in Hilbert
spaces $H_1$ and $H_2,$ respectively. Consider Cartesian product
of these Hilbert spaces: ${\bf H} = H_1 \times H_2$(don't mix
with $H = H_1 \otimes H_2)$ and the random vector   $\omega \to
\phi (\omega) = (\phi_1 (\omega), \phi_2 (\omega)) \in {\bf H}$
 such that: a)  its expectation $E \phi = 0;$ b) its dispersion
$\sigma^2 (\phi) = E ||\phi||^2 < \infty.$ Take its covariance
operator $D$ which is determined by the symmetric (positive)
bilinear form: $(D u, v) = E (u, \phi) (v, \phi),$ where vectors
$u, v \in {\bf H}.$ This operator has the block structure
$D = \left( \begin{array}{ll}
 D_{11} & D_{12}\\
D_{21} & D_{22}\\
 \end{array}
 \right ),
 $
 where $D_{ii}: H_i \to H_i, D_{ij}: H_j \to H_i.$

Let $\widehat{A}_i \in {\cal L}_s (H_i), i= 1,2.$ It determines
the quadratic function on the Hilbert space $H_i: f_{A_i}
(\phi_i) = (\widehat{A}_i \phi_i, \phi_i).$ Such quadratic functionals
are prequantum physical variables corresponding to quantum observables.

For any Gaussian random vector  $\phi = (\phi_1 , \phi_2 )$ having
zero average
 and any pair of operators
$\widehat{A}_i \in {\cal L}_s (H_i), i= 1,2,$ the following
equality takes place \cite{KHR}
 $$
  \langle f_{A_1}, f_{A_2}\rangle_\phi \equiv
 E f_{A_1}(\phi_1 ) f_{A_2} (\phi_2 )
$$
\begin{equation}
 \label{T00}
= ({\rm Tr} D_{11} \widehat{A}_1) ({\rm Tr} D_{22} \widehat{A}_2)
+ 2 {\rm Tr} D_{12} \widehat{A}_2 D_{21} \widehat{A}_1.
\end{equation}
We also remark that \cite{KHR} ${\rm Tr} D_{ii} \widehat{A}_i = E
f_{A_i} (\phi_i), i=1,2.$ Thus
 $E f_{A_1} f_{A_2} =  E f_{A_1} E f_{A_2} +
2 {\rm Tr} D_{12} \widehat{A}_2 D_{21} \widehat{A}_1.$
Now take an arbitrary pure state of a composite system $S=(S_1,
S_2),$ a normalized vector $\Psi \in H.$
Consider a Gaussian  vector random field such that $D_{12}=
\widehat{\Psi}.$  By operator equality (\ref{Zuzu}) the last summand
in the right-hand side of (\ref{T00}) is equal to the QM-average.
Hence, we obtain
 $\frac{1}{2} E (f_{A_1} - E f_{A_1}) (f_{A_2} - E f_{A_2}) =
(\widehat{A}_1 \otimes \widehat{A}_2 \Psi, \Psi) \equiv \langle
\widehat{A}_1 \otimes \widehat{A}_2 \rangle_\Psi,$
or, for the covariance of two classical random vectors $f_{A_1},
f_{A_2},$ we have:
\begin{equation}
 \label{Q1}
 \frac{1}{2} {\rm cov} \; (f_{A_1}, f_{A_2}) = \langle \widehat{A}_1 \otimes \widehat{A}_2 \rangle_\Psi.
\end{equation}
This formula was derived in \cite{KHR}. One of problems is its
asymmetry: classical covariance on one side, i.e., centered
correlation, but non-centered quantum correlation  on another
side. We shall obtain symmetric representation. However, first we
have to discuss a more fundamental problem

Operators $D_{ii}$ are responsible for averages of random
variables $\omega \to f (\phi_i (\omega)),$ i.e., depending only
on one of components of the vector random field $\phi.$ In
particular, $E f_{A_i} (\phi_i)  = {\rm Tr} D_{ii} \widehat{A}_i.$

We shall  construct such a random field that these averages will
match those given by QM. It is  natural to take the covariance
operator $\tilde{D}_\Psi= \left( \begin{array}{ll}
 \widehat{\Psi} \widehat{\Psi}^*  & \widehat{\Psi}\\
 \widehat{\Psi}^* & \widehat{\Psi}^*\Psi \;
 \end{array}
 \right ).
$ However, in general this operator is not positively defined and,
hence, it cannot serve as a covariance operator. In \cite{KHR} it
was proposed to modify aforementioned operator and consider
$D_\Psi= \left( \begin{array}{ll}
 \widehat{\Psi} \widehat{\Psi}^* + \epsilon I & \; \; \; \; \; \; \; \widehat{\Psi}\\
\; \;  \; \; \; \; \; \widehat{\Psi}^* & \widehat{\Psi}^*\Psi \;
+ \epsilon  I
 \end{array}
 \right ),
$ where $\epsilon >0$ is sufficiently large. We remark that white
noise is a Gaussian random variable with zero average and the unit
covariance operator $I.$ Thus additional terms in diagonal blocks
are related to the white noise background. The situation is
tricky: in general it is impossible (in the classical mathematical
model) to separate this noisy background from a random prequantum
field. We cannot consider a random field with the  covariance
operator $D_\Psi$ as the sum of two signals, e.g., an electron
signal and the background signal.  For some states (entangled
states), the matrix with $\epsilon =0$ is not positively defined.
We discuss this point in more detail:

Suppose now that $\phi (\omega)$ is a random vector with the
covariance operator $D_\Psi.$ Then
\begin{equation}
\label{YY1} \langle \widehat{A}_1 \rangle_\Psi = E f_{A_1} (\phi_1
(\omega)) - \epsilon {\rm Tr} \widehat{A}_1,
 \end{equation}
\begin{equation}
\label{YY2} \langle \widehat{A}_2 \rangle_\Psi = E f_{A_2} (\phi_2
(\omega)) - \epsilon {\rm Tr} \widehat{A}_2.
\end{equation}
These relations for averages and relation (\ref{Q1}) for the
correlation provide coupling between theory  of classical Gaussian
signals (in the finite-dimensional case simply theory of Gaussian
random variables) and QM. Quantum statistical quantities can be
obtained from corresponding quantities for classical random field.
One may say that {\it ``irreducible quantum randomness" is reduced
to randomness of classical prequantum fields.} However, the
situation is more complicated. The equalities (\ref{YY1}),
(\ref{YY2}) imply that quantum averages are obtained as the
shift-type renormalizations of averages with respect to classical
random fields. The shift corresponds to subtraction of the
contribution of the background field. Thus quantum averages are
not simply classical averages. They are obtained as the result of
renormalization with respect of the background field.

\section{Modification of correspondence between quantum observables and classical variables}
\label{MODF}

Although equality (\ref{Q1}) establishes the coupling between
classical correlations of random signals and quantum correlations,
it is not completely satisfactory from the purely probabilistic
viewpoint. On the left-hand side of (\ref{Q1}) we have the
classical {\it covariation}, $\rm {\rm cov} \; (f_{A_1},
f_{A_2})$, but on the right-hand side we have just the quantum
average of the correlation observable $\widehat{A}_1 \otimes
\widehat{A}_2.$ We want to modify the correspondence between
quantum and classical models to obtain a symmetric relation
between classical and quantum covariances. We recall that the
latter is given by ${\rm cov} \; (\widehat A_1, \widehat
A_2)\equiv$ $$\langle A_1 \otimes A_2 \rangle - \langle A_1
\rangle \langle A_2 \rangle = \langle (\widehat A_1 - \langle
\widehat A_1 \rangle I ) \otimes ( \widehat A_2 - \langle \widehat
A_2 \rangle I)\rangle .$$ We set
\begin{equation}
\label{QCOV1}
\widehat A_{0i} = \widehat A_i - \langle A \rangle I, i = 1,2,
\end{equation}
Then $\langle \widehat A_{0 i}\rangle = 0$ and
\begin{equation}
\label{QCOV2}
{\rm cov} \; (\widehat A_{01}, \widehat A_{02}) =
\langle \widehat A_{01} \otimes \widehat A_{02} \rangle = {\rm cov} \; (\widehat A_1, \widehat A_2).
\end{equation}

Let us modify the correspondence between classical and quantum variables, see section
\ref{EER}. Instead of the formerly used correspondence
$$\widehat A \to f_A(\phi) = ( \widehat A \phi, \phi),$$ we introduce a new map from the quantum model to the classical
prequantum model
\begin{equation}
\label{t7} \widehat A_0 \to f_{A_0}(\phi) = (\widehat A_0 \phi,
\phi ), \widehat A_0 = \widehat A - \langle A \rangle I.
\end{equation}
By using (\ref{Q1}) for $\widehat A_{0i}$ instead of $\widehat
A_i,$ we obtain
$${\rm cov}\; (f_{A_{01}}, f_{A_{02}}) = \langle \widehat A_{01} \otimes \widehat A_{02} \rangle
.$$
Thus
\begin{equation}
\label{t4} \frac{1}{2}{\rm cov}\; (f_{A_{01}}, f_{A_{02}}) = {\rm
cov} \; (\widehat A_{01}, \widehat A_{02}) .
\end{equation}
We remark that in (\ref{Q1}) the factor 2 in front of the quantum
correlation disappears in the complex case \cite{KHRJMP}. So, in
the complex case the correspondence becomes really symmetric. (We
proceed in the real Hilbert space, since in the complex case the
basis operator equality (\ref{Zuzu}) is more complicated and its
presentation is based on more complicated operator theory.)

\subsection{Independence of components of prequantum random fields corresponding
to factorizable quantum states}

Consider now a factorizable quantum state $\Psi = \Psi_1 \otimes \Psi_2,$ where
$\Psi_i \in H_i, i = 1,2.$ Then
$$\langle \widehat A_{01} \otimes \widehat A_{02}\rangle =
(\widehat A_{01} \Psi_1, \Psi_1 ) ( \widehat A_{02} \Psi_2, \Psi_2
) = 0.
$$
Hence,
\begin{equation}
\label{t3}
\rm{cov} \; (\widehat A_1, \widehat A_2) = \rm{cov} \; (\widehat A_{01}, \widehat A_{02}) = 0.
\end{equation}
Thus by (\ref{t4}),
$$
\rm{cov} \; (f_{A_{01}}, f_{A_{02}}) = 0.
$$

\medskip

Factorization of a pure quantum state $\Psi$ implies that, for any two quantum
observables $\widehat A_i$ on the subsystems $S_i, i = 1,2,$ of a composite system
$S = (S_1, S_2),$ the corresponding prequantum variables, $f_{A_{0i}} (\phi_i),$
where $\widehat A_{0i} = \widehat A_i - \langle A_i \rangle I,$ are uncorrelated.

\subsection{Totality of correlations at the subquantum level}

We remark that, although for a factorizable quantum state all
prequantum physical variables corresponding to quantum observables
are uncorrelated, see previous section, components
$\phi_1(\omega)$ and $\phi_2(\omega)$ of the prequantum field are
{\it always correlated.} The covariance operator $D_\Psi$ always
has nonzero off-diagonal block $D_{12}= \widehat{\Psi}.$ Thus we
can find other prequantum physical variables, nonquadratic
functionals of the prequantum field, which are nontrivially
correlated. Roughly speaking the presently used class of
observables is too restricted to find this totality of
correlations at the subquantum level. Not only prequantum fields
corresponding to entangled quantum systems are correlated, but
even prequantum fields corresponding to distinguishable quantum
systems which have been prepared independently at  huge distance
from each other. Thus {\it subquantum  nonlocality is even more
general  than quantum one.} However, the former is purely
classical nonlocality of correlations of random fields which are
coupled through the common background field.

\section{Distinguishing property of ``entangled prequantum
fields''}

Thus all prequantum random fields are correlated. Can one
distinguish random fields corresponding to entangled quantum
states from random fields corresponding to  factorizable states?
One of distinguishing features of ``entangled prequantum fields''
is impossibility to separate  the ``intrinsic field'' of a quantum
system from the background field. The intrinsic field does not
exist as classical random field. This is again the problem of
positive definiteness of the covariance operator. We remark that
the operator $\tilde{D}_\Psi= \left(
\begin{array}{ll}
 \widehat{\Psi} \widehat{\Psi}^*  & \widehat{\Psi}\\
\widehat{\Psi}^* & \widehat{\Psi}^*\Psi \;
 \end{array}
 \right )
$ is positively defined iff the quantum state $\widehat{\Psi}$ is
factorizable, $\Psi= \Psi_1\otimes \Psi_2.$ The step from
factorizability to positive definiteness is trivial. For
$\phi=(\phi_1, \phi_2) \in {\bf H},$ we have
$$
(\tilde{D}_\Psi \phi, \phi) = \Vert \widehat{\Psi}^*  \phi_1\Vert^2 + 2 (\widehat{\Psi}^* \phi_1, \phi_2) +
\Vert \widehat{\Psi}  \phi_2\Vert^2=
$$
$$
(\Psi_1, \phi_1)^2 + 2(\Psi_1, \phi_1) (\Psi_2, \phi_2) + (\Psi_2, \phi_2)^2 \geq 0.
$$
Suppose now that $\Psi$ is not factorizable. Consider its Schmidt
decomposition $\Psi = \sum_i \alpha_i e_{i1}\otimes e_{i2},$ where
$\{e_{i1}\}$ and $\{e_{i2}\}$ are orthonormal systems in $H_1$ and
$H_2,$ respectively. We shall explore the following property of
Schmidt decomposition: it contains just one summand if and only if
$\Psi$ is factorizable. Consider coordinates $x_i=(\phi_1,
e_{i1}), y_i= (\phi_2, e_{i2}).$ Then $(\tilde{D}_\Psi \phi, \phi)
=\sum_i (\alpha_i^2 x_i^2 +  2 \alpha_i x_i y_i + \alpha_i^2
y_i^2).$ In the case of entanglement all coefficients $\alpha_i
<1.$ Set $x_i=y_i=0,$ for  $i>1.$ Then $(\tilde{D}_\Psi \phi,
\phi) = \alpha_1^2 x_1^2 + 2 \alpha_1 x_1 y_1 + \alpha_1^2 y_1^2.$
This quadratic form is not positively defined.

In the case of a prequantum random field corresponding to a
factorizable quantum state the operator $D_\Psi= \tilde{D}_\Psi +
\epsilon I$ and the first summand is positively defined. Consider
Gaussian random fields $\phi_\Psi,  \tilde{\phi}_\Psi, \eta$ with
zero mean values and the covariance operators $D_\Psi,
\tilde{D}_\Psi , \epsilon I.$ Suppose that the ``intrinsic field''
of a system  $\tilde{\phi}_\Psi$ and the background field $\eta$
are independent.  Then $\phi_\Psi$ can be represented as
$\phi_\Psi(\omega)= \tilde{\phi}_\Psi(\omega) +\eta(\omega),$
where $\omega$ is a random parameter. Thus in the absence of
entanglement the ``intrinsic field'' of a system can be distilled
from the background.  This is impossible for entangled states.

\medskip

{\bf Conclusion.} {\it At the subquantum level entanglement is an
exhibition of fundamental nonseparability from the background
field.}

\end{document}